\documentclass{jfm}
\usepackage{graphicx}
\usepackage{amssymb}
\usepackage{natbib}
\begin{document}

\title{Length and time scales of a liquid drop impact and penetration into a granular layer} 

\author{Hiroaki Katsuragi}
\affiliation{Department of Applied Science for Electronics and Materials, Kyushu University, 6-1 Kasugakoen, Kasuga, Fukuoka 816-8580, Japan}

\begin{abstract}
Liquid drop impact and penetration into a granular layer are investigated with diverse liquids and granular materials. We use various size of SiC abrasives and glass beads as a target granular material. We also employ ethanol and glycerol aqueous solutions as well as distilled water to make a liquid drop. The liquid drop impacts the granular layer with a low speed ($\sim$m/s). The drop deformation and penetration are captured by a high speed camera. From the video data, characteristic time scales are measured. Using a laser profilometry system, resultant crater morphology and its characteristic length scales are measured. Static strength of the granular layer is also measured by the slow pillar penetration experiment to quantify the cohesive force effect. We find that the time scales are almost independent of impact speed, but they depend on liquid drop viscosity. Particularly, the penetration time is proportional to the square root of the liquid drop viscosity. Contrastively, the crater radius is independent of the liquid drop viscosity. The crater radius is scaled by the same form as the previous paper, (Katsuragi, Phys. Rev. Lett. vol.~104, 2010, p. 218001).
\end{abstract}


\date{\today}

\maketitle

\section{Introduction}\label{sec:introduction}
Granular system behaves sometimes like fluid, but sometimes like solid~(\cite{Jaeger1996,Duran2000,Aranson2006}). If a collisional impulse is applied to a granular matter, the granular matter behaves like fluid at the impact moment. After that, it relaxes right away and behaves like solid at a later stage. Thus the granular response observation against the impact can prove both of fluid-like and solid-like behaviors. In this sense, the granular impact experiment is very useful to study the granular matter dynamics. A lot of granular impact experiments using a solid projectile have been carried out recently. Some of them were interested in the crater morphology~(\cite{Amato1998,Uehara2003,Walsh2003,Ambroso2005_1,deVet2007}), and others discussed the dynamics~(\cite{Ciamarra2004,Ambroso2005_2,Hou2005,Katsuragi2007,Royer2007,Nelson2008,Goldman2008}). Granular jet~(\cite{Thoroddsen2001,Lohse2004N,Lohse2004P,Bergmann2005,Royer2005,Caballero2007,vonKann2010} and ejecta~(\cite{Boudet2006,Deboeuf2009}) created by the impact have been also studied. Numerical studies have been performed as well~(\cite{Tsimring2005,Wada2006}). Thanks to these numerous studies, granular impact morphology and dynamics are partly understood. However, all these results have concerned the solid projectile impact. For the solid projectile impacts, the projectile is much harder than the target granular layer. There was no need to consider the projectile deformation effect. 
Then the next natural question, which is the main focus of this paper, is "what happens when the projectile is deformable?" In order to answer this question, we have performed the impact experiment of liquid drop and granular layer. 

A liquid drop impact to a hard wall or a liquid pool phenomenon has been also studied extensively~(\cite{Marmanis1996,Range1998,Throddsen1998,Bhola1999,Rioboo2002,Yarin2006,Xu2010}). The milk-crown shape by the drop impact is very beautiful and fascinating. Additionally, this drop impact spreading and splashing phenomena relate to many industrial technologies, such as ink-jet printing, rapid spray cooling, mixing of solutions, and so on~(\cite{Yarin2006}). Although the slow penetration of liquid drops into a powder bed has been reported~(\cite{Hapgood2002}), the impact between liquid drop and granular layer has not been examined until very recently~(\cite{Delon2009,Katsuragi2010_1,Nefzaoui2010}). Dalziel and Seaton have studied similar phenomenon of the impact between a droplet and a granular layer. But their result was very limited and focused only on resuspension effect~(\cite{Dalziel2003}).

In the previous study~(\cite{Katsuragi2010_1}), SiC abrasives and water drops were used to characterize typical crater shapes and to reveal the crater radius scaling. The impact experiments with glass beads and some kinds of drops have been carried out recently by another research group~(\cite{Nefzaoui2010}). They have derived scalings for drop deformation and crater diameter. However, these experiments were still limited to reveal the overall characteristics of the drop-granular-impact. In order to clarify the details of the drop-granular-impact, we have performed experiments with various liquid drops and granular layers.

In laboratory experiments, parameter range we can access is actually very limited. For instance, it is hard to make a large liquid drop since the capillary length of usual liquids on the earth surface is about mm scale~(\cite{deGennes2004}). In our experiment, the impact speed made by the free fall is also limited within the order of m/s. While the experimental conditions are limited as written above, the drop-granular-impact experiment can be a key to understand the geological scale events. Actually, low speed impact of small solid impactor to very loose granular layer resembles large planetary impacts, in terms of the dimensionless similarity law~(\cite{Bergmann2005}). In addition, real geological impactors in the space might be fully broken when it hits the other astronomical objects~(\cite{Melosh1989}). Furthermore, the gravitational force might be smaller than the earth surface. Large liquid drops are possible in such environment. Even if the drop size is limited in mm scale, the drop-granular-impact might relate to the fossil rain drops. The fossil rain drops are the small pit-like depressions found in fine grain sediment~(\cite{Desor1850,Metz1981}). The origin of this fossil shape is still a matter of debate. The morphological study of the drop-granular-impact might give a hint to understand the origin of the fossil shape.

In the next section, the experimental apparatus, materials and procedures are introduced. In section~\ref{sec:dynamic_morphology}, drop deformation and impact dynamics are discussed based on high speed video data. In section~\ref{sec:static_strength_measurement}, a static strength measurement method is introduced to characterize the strength of the granular layer's surface. We discuss the drop deposition behavior observed in small glass beads impacts, using this measurement result. Next, the crater morphology is studied based on three dimensional (3D) surface measurement data~(section~\ref{sec:crater_morphology}). After that, typical time scales are measured and discussed using the video data~(section~\ref{sec:time_scale_analysis}). In particular, we focus on the viscosity dependence of the time scales. Finally, characteristic length scales of resultant crater shapes are measured and discussed~(section~\ref{sec:length_scale_analysis}). We find the simple approximate relation between crater depth and intersectional area. Previously proposed scaling for the crater radius is verified for various experimental conditions. Discussion and conclusion are stated in sections~\ref{sec:discussion} and \ref{sec:conclusion} .

\section{Experimental}\label{sec:experimental}
\begin{figure}
\begin{center}
\includegraphics[width=7.5cm]{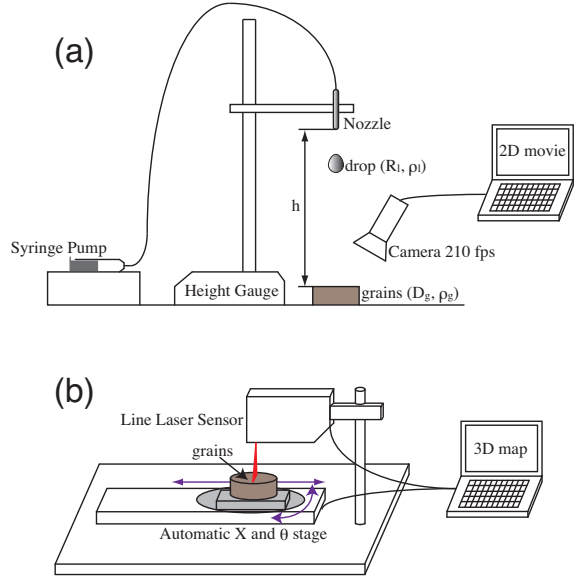}
\caption{Schematic image of the experimental apparatus. (a) A liquid drop is released from a nozzle tip mounted on a height gauge. The impact is captured by a high speed camera. (b) The resultant crater shape is measured by the laser profilometry system composed by a line laser sensor and automatic stages.}
\label{fig:apparatus}
\end{center}
\end{figure}

Experimental apparatus we built is shown in figure~\ref{fig:apparatus}. A nozzle is mounted on a height gauge. The granular target is set below it. Using a syringe pump, liquid is delivered slowly ($0.2$~ml/min.) to the nozzle. A drop grows gradually on the nozzle's tip, and is finally pinched off by the gravitational force. Then the drop commences the free fall. We vary the nozzle tip inner diameter ($0.25$, $0.7$, $2.4$, or $2.5$~mm) to control the drop size. Radius of the drop $R_l$ made by this procedure ranges from $R_l = 1.3$ to $2.4$~mm. The largest value ($2.4$~mm) is almost the upper limit for a drop under the gravity. We define the free fall height $h$ as height difference between the nozzle tip and the surface of the granular layer. It is adjusted by the height gauge. The drop elongates vertically up to $\sim 8$~mm just before its break. Thus the minimum value of $h$ is chosen as $10$~mm, and the maximum value of $h$ is $480$~mm. Impact speed $v$ is calculated from the free fall relation $v=\sqrt{2gh}$, where $g$ is the gravitational acceleration~($g=9.8$~m/s$^2$). Air drag deceleration is negligible in such low speed regime~(\cite{Range1998}). 

Impacts are captured by a high speed camera, TAKEX FC350CL, which takes $210$~fps video data with $640 \times 480$ pixels in $8$~bit grayscale images. After drop penetration, the crater shape is measured by a line laser displacement sensor based profilometry system. This system consists of a line laser displacement sensor, KEYENCE LJG030, and automatic stages, COMS PM80B-100X (translational motion) and COMS PS60BB-360R (rotational motion). An intersectional profile is obtained by the line laser sensor at a certain position. Then the sample is shifted using the stage, and the profile is obtained at that position. This procedure is repeated. Finally, the intersectional profiles at different positions are combined to make a 3D surface depth map. The system has $33$ and $1$~$\mu$m horizontal and vertical resolutions, respectively. Almost the same system has been applied to measuring the polymer gel surface pattern formation~(\cite{Mizoue2010}).  Our previous study of the drop-granular-impact has also used the same system~(\cite{Katsuragi2010_1}). Data handling and all instruments control are conducted by LabVIEW software. 

\begin{figure}
\begin{center}
\includegraphics[width=7.5cm]{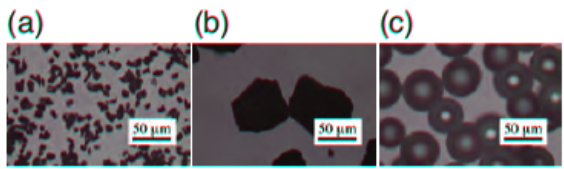}
\caption{Photos of grains: SiC abrasives of $D_g$ = (a) $4$ and (b) $50$~$\mu$m, and glass beads of $D_g=$  (c) $50$~$\mu$m. SiC grains are nonspherical and possess polydispersity, while glass beads are spherical and little polydisperse.
}
\label{fig:grains}
\end{center}
\end{figure}

We use SiC abrasives and glass beads for granular targets. The material true densities are $3.2$ and $2.5$~g/cm$^3$ for SiC and glass beads, respectively. Microscopic photos of SiC grains, which reveal that the SiC grains possess nonspherical shape and polydispersity, are shown in figures~\ref{fig:grains}(a,b). The central value of the distributed grain diameter $D_g$ is (a) 4 and (b) 50~$\mu$m. Glass beads ($D_g=50$~$\mu$m) image is shown in figure~\ref{fig:grains}(c). Glass beads possess good spherical shape and small polydispersity. We employ $D_g=4$, $8$, $14$, $20$, and $50$~$\mu$m SiC grains~(supplied by Showa denko k.k.; JIS R6001 \#3000, \#1500, \#800, \#600, and \#220, respectively) and $D_g=5$, $18$, $50$, and $100$~$\mu$m glass beads~(supplied by Potters-Ballotini Co., Ltd.). A small vessel ($30$~mm in diameter and $10$~mm thick) is filled with grains as a target. Filling and surface flattening of piled grains are performed by hand. Grains are piled up on the vessel using a spoon, and then a flat metal plate sweeps the pile to make a smooth flat surface. No additional tapping or other compacting process is applied. This simple preparation method makes good reproducibility. The uncertainty of the bulk density is a few percent. The impact crater shape also shows good reproducibility. Granular properties we use are summarized in table~\ref{tab:grains}. All packing fraction values have less than $0.02$ uncertainty. SiC grain's irregular shape and polydispersity result in a very small packing fraction~(\cite{Roller1930,Suzuki2001}). Even for spherical glass beads, small packing fraction is observed. In such case, grains are too small, and cohesive force affects the packing state. This cohesive small glass beads layer is very strong against the impact, as described later. In addition, the surface roughness of the granular layer brings a lotus effect, which affects the fragility of the impacted drop~(\cite{Reyssat2008}). 

\begin{table}
\begin{center}
\begin{tabular}{lccc}
{Material} & {$D_g$~($\mu$m)} & {$\rho_g$~(g/cm$^3$)} & {packing fraction} \\
SiC abrasives & $4$ & $0.99$ & $0.31$ \\
SiC abrasives & $8$ & $0.99$ & $0.31$ \\
SiC abrasives & $14$ & $1.5$ & $0.44$ \\
SiC abrasives & $20$ & $1.5$ & $0.44$ \\
SiC abrasives & $50$ & $1.5$ & $0.50$ \\
Glass beads & $5$ & $1.0$ & $0.39$ \\
Glass beads & $18$ & $1.4$ & $0.55$ \\
Glass beads & $50$ & $1.5$ & $0.59$ \\
Glass beads & $100$ & $1.6$ & $0.63$ \\
\end{tabular}
\end{center}
\caption{Granular layer properties: grain size $D_g$, bulk density~$\rho_g$, and packing fraction of target granular layer.}
\label{tab:grains}
\end{table}

We also varied liquid drop surface tension and viscosity as well as drop size. Ethanol and glycerol are utilized to control surface tension and viscosity. Distilled water, $10$wt\% and $100$wt\% ethanol aqueous solutions, and $60$wt\%, $88$wt\% and $100$wt\% glycerol aqueous solutions are employed to examine the surface tension and viscosity effect. Liquid properties are listed in table~\ref{tab:fluids}. These values are picked up from tables in the data book~(\cite{JSME1983}).

\begin{table}
\begin{center}
\begin{tabular}{lcccc}
{Liquid} & {$R_l$ (mm)} & {$\rho_l$ (g/cm$^3$)} & {$\gamma$ (mN/m)} & {$\nu$ (10$^6$ m$^2$/s)} \\
distilled water  & $1.3$ - $2.4$ & $1.0$ & $72$ & $0.89$ \\
ethanol 10wt\% & $1.9$ & $0.98$ & $47$ & $1.5$ \\
ethanol 100wt\% & $1.5$ & $0.79$ & $22$ & $1.5$ \\
glycerol 60wt\% & $2.0$ & $1.2$ & $68$ & $9.6$ \\
glycerol 88wt\% & $2.0$ & $1.2$ & $65$ & $120$ \\
glycerol 100wt\% & $1.9$ & $1.3$ & $63$ & $1400$ \\
\end{tabular}
\end{center}
\caption{Liquid drop properties~(\cite{JSME1983}). $R_l$, $\rho_l$, $\gamma$, and $\nu$ are drop radius, density, surface tension, and kinematic viscosity, respectively.
}
\label{tab:fluids}
\end{table}

\section{Results}\label{sec:results}

\subsection{Dynamic morphology}\label{sec:dynamic_morphology}
\begin{figure}
\begin{center}
\includegraphics[width=13cm]{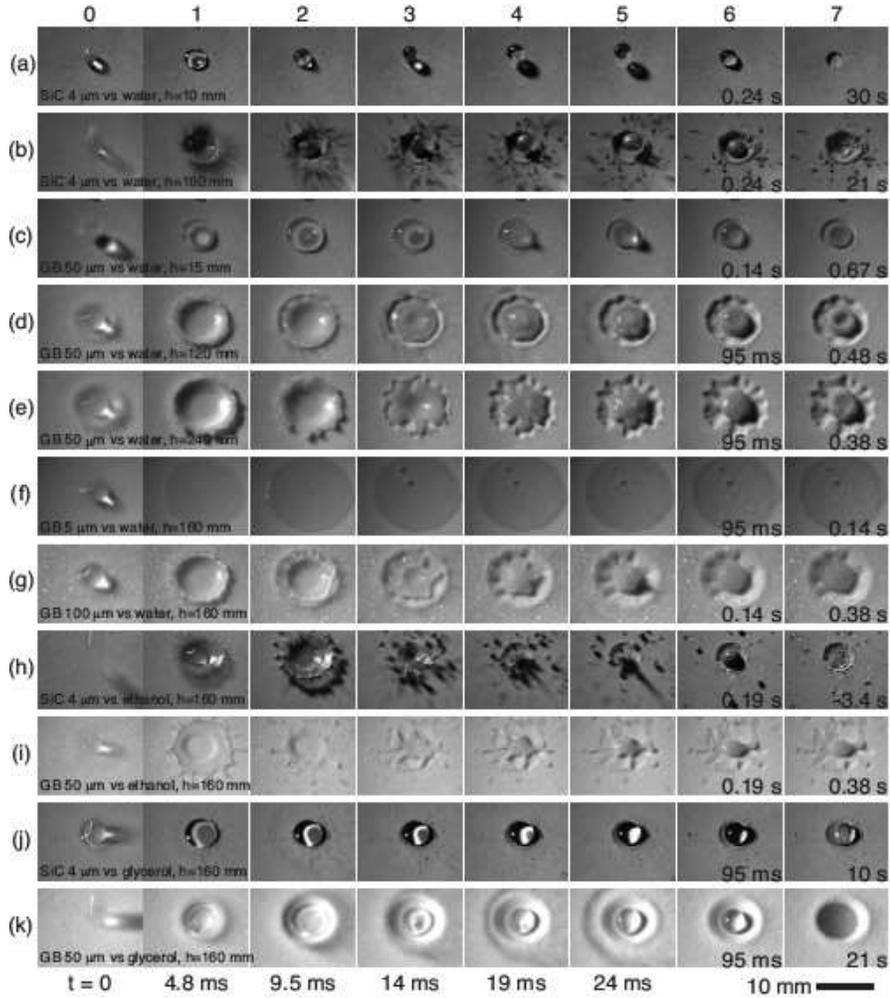}
\caption{Photos of raw data taken by a high speed camera. Each row corresponds to each experimental condition as cited on each left bottom part. Specific experimental conditions are as follows: (a) SiC of $D_g=4$~$\mu$m vs water drop of $R_l=1.3$~mm with $h=10$~mm, (b) same as (a) except $h=160$~mm, (c) glass beads of $D_g=50$~$\mu$m vs water drop of $R_l=2.2$~mm with $h=15$~mm, (d) same as (c) except $h=120$~mm, (e) same as (c) except $h=240$~mm, (f) glass beads of $D_g=5$~$\mu$m vs water drop of $R_l=2.2$~mm with $h=160$~mm, (g) glass beads of $D_g=100$~$\mu$m vs water drop of $R_l=2.2$~mm with $h=160$~mm, (h) SiC of $D_g=4$~$\mu$m vs pure ethanol (100wt\%) drop of $R_l=1.5$~mm with $h=160$~mm, (i) glass beads of $D_g=50$~$\mu$m vs pure ethanol (100wt\%) drop of $R_l=1.5$~mm with $h=160$~mm, (j) SiC of $D_g=4$~$\mu$m vs pure glycerol ($100$wt\%) drop of $R_l=1.9$~mm with $h=160$~mm, and (k) glass beads of $D_g=50$~$\mu$m vs pure glycerol (100wt\%) drop of $R_l=1.9$~mm with $h=160$~mm. Column $0$ corresponds to the impact moment~($t=0$). Column $1$ to $5$ display successive images right after the impact ($1/210$ s interval). Column $6$ and $7$ show later penetration stages. Scale bar ($10$~mm) is shown at the right bottom. As can be seen, various drop deformation and crater morphology are observed.
}
\label{fig:photos}
\end{center}
\end{figure}

First, we focus on high speed video data. In figure~\ref{fig:photos}, snap shots of various impacts are presented. Each row corresponds to different experimental condition. Column $0$ indicates the impact moment. It is the definition of the origin of time, $t=0$. The impact moment is identified by the drop deformation in the video data. This identification method is not very precise. The frame rate of the video data is not also very fast. The accuracy is almost determined by the frame rate as $1/210$ s. Column 1-5 represent the successive photos with 210 fps, right after the impact. Column 6 and 7 show the final penetration stage.

In figures~\ref{fig:photos}(a,b), small water drop impacts are shown. The water drop radius is $R_l=1.3$~mm, and the free fall height $h$ is (a) $10$~mm and (b) $160$~mm. One can confirm (a) the sink-type crater and (b) the ring-type crater. They have been reported in our previous paper with large drop size ($R_l=2.4$~mm)~(\cite{Katsuragi2010_1}). The drop deformation dynamics are qualitatively similar to the previous result, as well as the final crater shapes. This implies that the drop size hardly affects the drop deformation and crater morphology, as long as the drop size is in mm order. Much smaller drops of the $\mu$m or nm size may reveal completely different impact dynamics and crater morphology, though. The drop penetration in these processes is really gentle. They never show any sudden change. This process is very contrastive to the fakir-impale transition of the drop on a rough hydrophobic surface~(\cite{Reyssat2008}).

Typical pictures of the glass beads impacts are shown in figures~\ref{fig:photos}(c-e). In all situations, inner ring structure can be observed. Even in slow impact case~(figure~\ref{fig:photos}(c)), sink-type crater cannot be obtained. Another characteristic feature of the glass beads impact is an enhancement of the fingering instability. We can confirm the fingering instability and its remnant crater rim shape~(figures~\ref{fig:photos}(d,e)). This fingering instability can be often observed for $D_g \geq 250$ $\mu$m cases. This result is natural since the surface roughness of the granular layer should influence the spreading of the drop~(\cite{Reyssat2008}).

When the grain size of glass beads is too small, the drop behavior becomes quite different. We show the $D_g=5$~$\mu$m glass beads impact in figure~\ref{fig:photos}(f). We are not able to see any cratering in this situation. Instead, the water drop spreads on the surface of glass beads layer. And it penetrates into the layer very rapidly. Actually, we observe this deposition behavior in all free fall heights ($h=10$ - $480$~mm) for $D_g=5$~$\mu$m glass beads. We think this deposition occurs due to the cohesive force among glass beads. The principal origin of this cohesive force is capillary bridge effect which is emphasized in smaller particles under the laboratory humidity ($50 \pm 10$\%) condition. Since glass is very hydrophilic material, the capillary cohesion makes a very strong glass beads layer. Hydrophilic glass surface also helps rapid penetration. As a consequence, drop deposition takes place in this situation. It should be noticed that small SiC grains~($D_g=4$~$\mu$m) are able to make various crater shapes, i.e., deposition never happens. This means that the SiC grains are hydrophobic compared to glass beads. This consideration is supported with the time scale analysis discussed later. Hydrophilicity of target grains affects the impact and penetration dynamics. To characterize this effect, we performed another slow penetration experiment. A steel pillar was penetrated into a granular layer very slowly. The result of this experiment is discussed in the next section. We have actually observed the cohesive effect in $D_g=18$~$\mu$m glass beads as well. In that case, very shallow rim and very small central ring or bump were made. These shapes are too thin to be measured using our measurement system. Thus we remove these cases (glass beads $D_g=5$ and $18$~$\mu$m) from the characteristic time and length scales analyses below. 

In figure~\ref{fig:photos}(g), large glass beads ($D_g=100$~$\mu$m) case is displayed. The drop deformation and outer rim structure are similar to those of figure~\ref{fig:photos}(d). The resultant crater inner shape is close to the bump-type (convex one) rather than the ring-type (concave one). This trend (bump-type crater appears in large $D_g$ grains impact) is same as the SiC result~(\cite{Katsuragi2010_1}). 

Next, we show the low surface tension liquid drop cases in figures~\ref{fig:photos}(h,i). A pure ethanol (100wt\%) drop impacts to (h) SiC of $D_g=4$~$\mu$m and (i) glass beads of $D_g=50$~$\mu$m. As can be seen, the drop is no longer stable owing to the fingering instability. Impact inertia easily fragments the drop to smaller drops. In such case, resultant crater size evidently becomes smaller than stable drop case. Moreover, unstable drop structure is sometimes frozen to the irregular shaped bump, as shown in fgiure~\ref{fig:photos}(i). Because we would like to focus only on the drop without fragmentation, we remove these data (ethanol 100wt\% impacts) from the characteristic time and length scales analyses. Ethanol 10wt\% impacts are more or less stable. We use only the ethanol 10wt\% impact data for the analyses below. 

Finally, viscous drop impacts are shown in figures~\ref{fig:photos}(j,k). A pure glycerol (100wt\%) drop impacts to (j) SiC of $D_g=4$~$\mu$m and (k) glass beads of $D_g=50$~$\mu$m. Since the viscosity of glycerol is three orders of magnitude greater than water (table~\ref{tab:fluids}), the impact dynamics is quite different. The most significant difference is deformability of the drop. The glycerol drop deforms itself much less than the water drop (compare, e.g., fgiures~\ref{fig:photos}(d,g, and k)). Another characteristic feature of viscous impact is slow penetration. Since the drop penetration takes very long time, we cannot obtain the final crater shape using high speed video data as shown in figure~\ref{fig:photos}(j). We measured the penetration time using another slow video camera in such case. And we will discuss the viscosity dependence of the penetration time scale in section~\ref{sec:time_scale_analysis}.

\subsection{Static strength measurement}\label{sec:static_strength_measurement}
\begin{figure}
\begin{center}
\includegraphics[width=7.5cm]{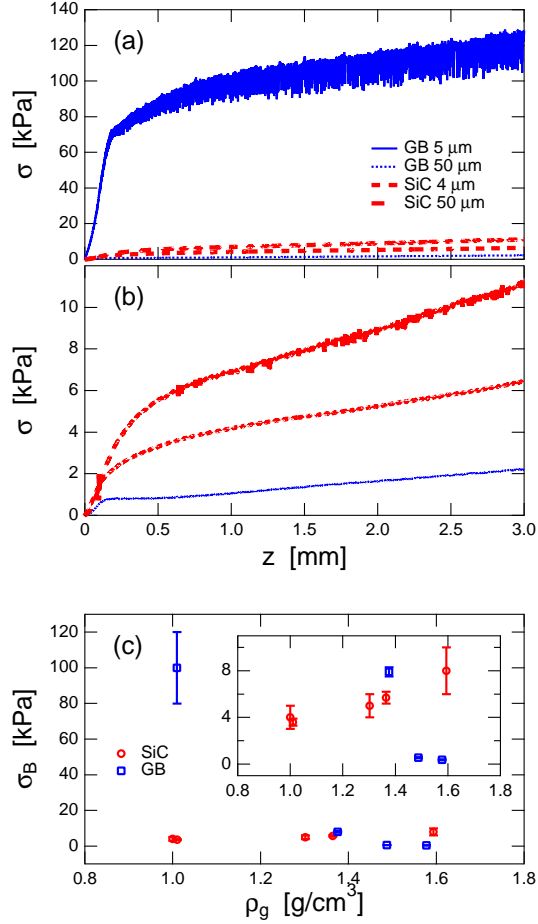}
\caption{Static stress measurement results. A steel pillar ($5$~mm diameter) was slowly penetrated into a granular layer, and the stroke $z$ and the stress $\sigma$ applied to the pillar were measured. (a) Stress vs stroke for various target grains are shown. (b) is the bottom part of (a). (c) Breaking stress $\sigma_B$ as a function of granular bulk density $\rho_g$. $\sigma_B$ denotes the crossover between elastic and hydrostatic behaviors. The inset shows the blow up of the bottom part.
}
\label{fig:sigmaB}
\end{center}
\end{figure}

As mentioned above, small glass beads layer is very strong due to the capillary bridge effect. To characterize this effect, we measured the static strength of the granular layer using a universal testing machine (Shimadzu AG100NX). A steel pillar with $5$~mm diameter was penetrated into the granular layer very slowly ($0.1$~mm/min.). The granular samples we used here were completely same as the impact experiment ones. Stroke $z$ and stress $\sigma$ applied to the pillar were measured using the testing machine. The obtained stroke-stress relations are shown in fiures~\ref{fig:sigmaB}(a,b). In all $\sigma$ data, crossover behavior is confirmed. At large $z$, asymptotic linear relation is observed. We think this comes from the hydrostatic pressure of the granular layer~(\cite{Stone2004a,Stone2004b}). This hydrostatic regime lasts to $z=3$ mm. This means that the bottom wall effect is negligible up to this depth. All the crater depth is shallower than this value. On the other hand, steeper linear relation appears in very shallow regime~($z<0.5$~mm). We assume this is a kind of elastic response of the granular layer. The origin of this pseudo elastic property is the grains cohesive force due to the capillary bridge effect among grains. This elastic regime and its limit stress are actually very small.

Here, we define the crossover stress value as the breaking stress $\sigma_B$. We first compute the linear fitting for large $z$ regime. And, $\sigma_B$ is determined by the $\sigma$ value at which the data deviate from the fitting. In figure~\ref{fig:sigmaB}(c), $\sigma_B$ is plotted as a function of granular bulk density $\rho_g$. Each data point is obtained by the average of three independent runs. As expected, the $\sigma_B$ of $D_g=5$~$\mu$m glass beads layer is much greater than others. This implies that the small glass beads layer is very strong against the stress. Liquid drop impact cannot deform such strong granular layer. That is a reason why the drop deposition is observed in small glass beads impacts (figure~\ref{fig:photos}(f)). The $\sigma_B$ of $D_g=18$~$\mu$m glass beads is about $8$ kPa. SiC of $D_g=50$~$\mu$m also shows $\sigma_B \simeq 8$ kPa. These cases are marginal ones to make various crater shapes or not, by the drop impact. Therefore, we consider $\sigma_B = 8$ kPa is the critical value for our experiment. If the granular surface strength $\sigma_B$ is greater than $8$ kPa, the granular surface is hardly deformed by the liquid drop impact. The estimated impact inertia stress by $\rho_l=1 \times 10^{3}$~kg/m$^3$ and $v=3$~m/s ($\sim h=480$~mm) is $\rho_l v^2 = 9$ kPa. Where $\rho_l$ is density of the liquid drop. This value is close to the measured marginal stress $8$ kPa. This critical stress also coincides with the single drop stability limit. Around $h=480$~mm, water drop fragmentation begins to take place, particularly for larger grains cases ($D_g \geq 50$~$\mu$m). We can reduce this capillary bridge effect by baking glass beads. But it was hard to completely get rid of this effect under our laboratory environment. Except for the cohesive regime (glass beads of $D_g \leq 18$~$\mu$m), $\sigma_B$ of SiC is greater than that of glass beads (see inset of figure~\ref{fig:sigmaB}(c)). We guess this is due to the grains shape effect. Since SiC grains have irregular shape, the grains network connections are stronger than spherical grains. 

Note that this measurement is a kind of quasi static one. We assume this sort of measurement is relevant to discuss the granular strength against the impact. In general, static and dynamic properties of granular system do not have to agree quantitatively. Nonetheless, dynamic characterization using the static property works well in this case.

\subsection{Crater morphology}\label{sec:crater_morphology}
After the penetration of liquid drop, final crater shape was measured by the laser profilometry system. In this section, we discuss the morphology of the crater. 

\begin{figure}
\begin{center}
\includegraphics[width=7.5cm]{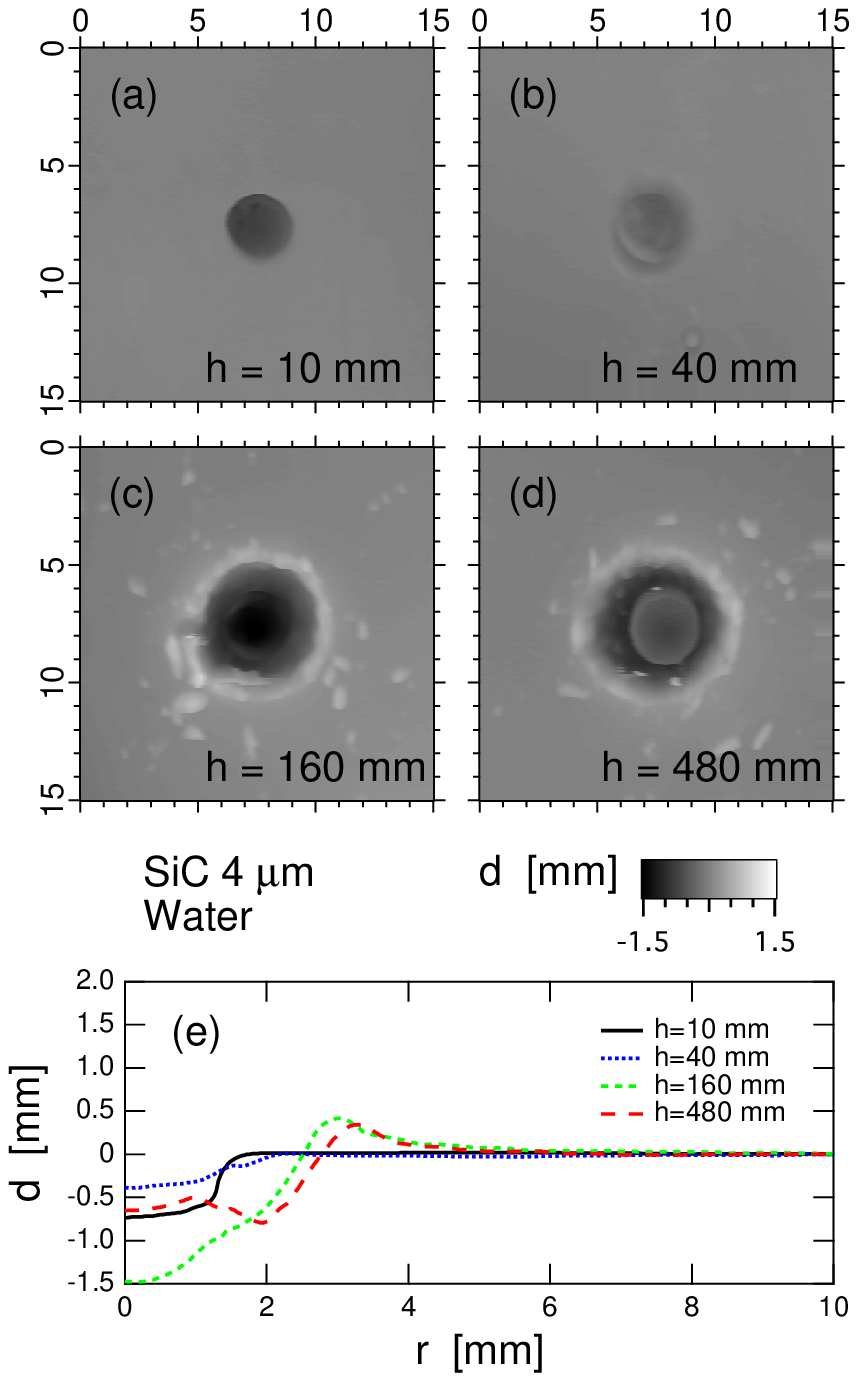}
\caption{Surface data of SiC grains ($D_g=4$~$\mu$m) vs small water drop ($R_l=1.3$~mm). (a-d) Surface height maps are shown as gray scale images. (e) Corresponding radial depth functions $d(r)$ are computed from raw height maps (a-d). Qualitative crater structure is similar to the large drop case. Sink-type (a) and ring-type (b-d) craters can be observed.
}
\label{fig:S1}
\end{center}
\end{figure}

First, craters by the small water drop vs small SiC grains layer ($D_g=4$~$\mu$m) are shown in figure~\ref{fig:S1}. Figures~\ref{fig:S1}(a-d) show raw 3D surface depth maps of different free fall heights. The depth $d$ is denoted as gray scale. Corresponding radial depth functions $d(r)$ are computed using these surface depth maps, and shown in figure~\ref{fig:S1}(e). Where $r$ is distance from the center of crater. One can confirm the deep sink-type crater at low speed impact. And, clear ring-type crater is observed in high speed impact. This tendency is completely same as large water drop impact case reported by~(\cite{Katsuragi2010_1}). We do not see any qualitative difference between large and small water drop impacts, at least under our experimental conditions.  

\begin{figure}
\begin{center}
\includegraphics[width=7.5cm]{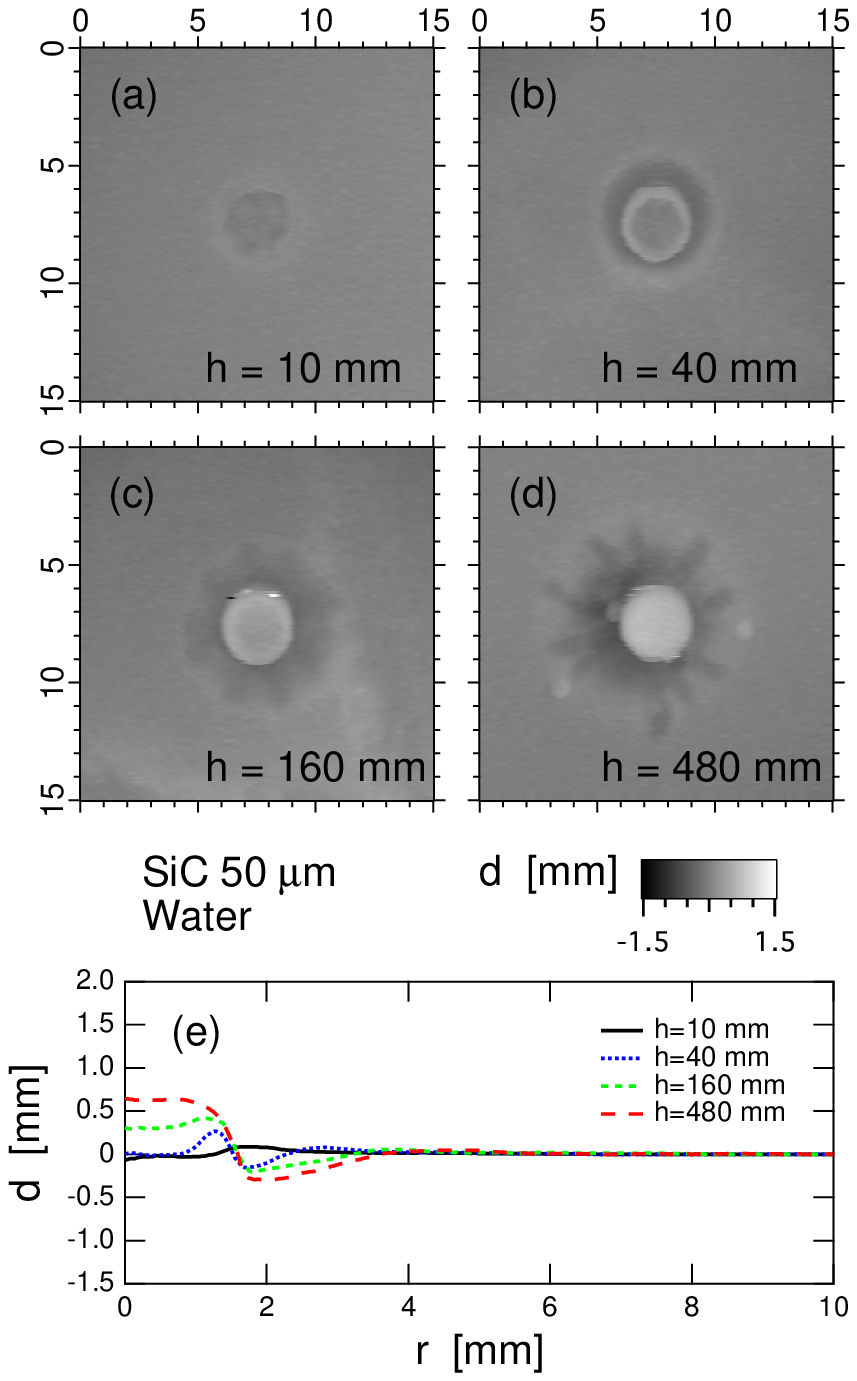}
\caption{Surface data of SiC grains ($D_g=50$~$\mu$m) vs small water drop ($R_l=1.3$~mm). (a-d) Surface height maps are shown as gray scale images. (e) Corresponding radial depth functions $d(r)$ are computed from raw height maps (a-d). Qualitative crater structure is similar to the large drop case again. Shallow sink-type (a), ring-type (b,c) and bump-type (d) craters can be observed.
}
\label{fig:S2}
\end{center}
\end{figure}

Next, the large SiC grains ($D_g=50$~$\mu$m) and small water drop ($R_l=1.3$~mm) impacts are displayed in figure~\ref{fig:S2}. The format of the figure is identical to figure~\ref{fig:S1}. Very shallow sink-type crater is observed at low speed impact, and bump-type crater is seen at high speed impact. This trend is same as large water drop impacts~(\cite{Katsuragi2010_1}) again. The only difference between small and large water drop is absolute crater size. For the characteristic length scale analysis in section~\ref{sec:length_scale_analysis}, we use the crater radius normalized to the original drop radius.

\begin{figure}
\begin{center}
\includegraphics[width=7.5cm]{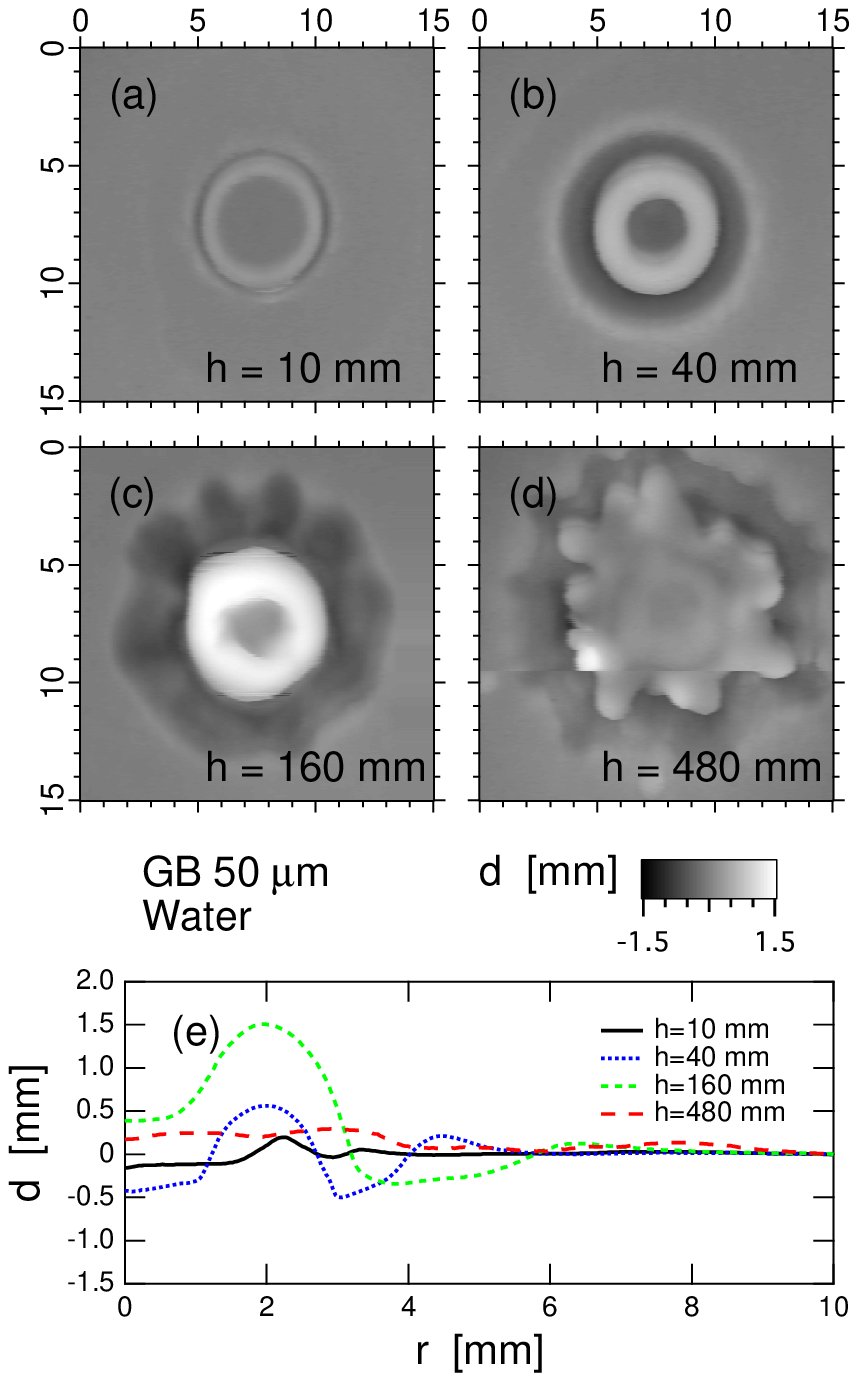}
\caption{Surface data of glass beads ($D_g=50$~$\mu$m) vs water drop ($R_l=2.2$~mm). (a-d) Surface height maps are shown as gray scale images. (e) Corresponding radial depth functions $d(r)$ are computed from raw height maps (a-d). Clear ring-type structure appears in a broad range of free fall height~($h \leq 160$~mm). At $h=160$~mm (c), fingering instability of the drop is recorded in the crater rim structure. At the highest free fall height $h=480$~mm (d), fingering instability affects the shape of the inner bump. 
}
\label{fig:S3}
\end{center}
\end{figure}

In fiure~\ref{fig:S3}, water drop ($R_l=2.2$~mm) and glass beads layer ($D_g=50$~$\mu$m) impacts are shown. The crater morphology of glass beads case is a little bit different from SiC grains case. Even in low impact speed regime, clear ring-type crater is observed. While the crater depth is not very deep ($d(r)$ is always greater than $-0.5$~mm), inner ring's rim height is much higher than SiC impacts. At $h=160$~mm, the maximum height is about $1.5$~mm. Moreover, the depth at the center $d(0)$ is greater than $0$. Thus this ring-type crater has a feature of the bump-type crater as well. The remnant of the drop fingering instability can be observed in figure~\ref{fig:S3}(c). We observe the asymmetric crater shape at the fastest impact case~(figure~\ref{fig:S3}(d)). This irregular shape also comes from the fingering instability. Since the glass is very hydrophilic material, water drop can penetrate into the glass beads layer faster than recovery of the spherical shape by the surface tension. Although the shape is very irregular, corresponding $d(r)$ is computed from the average of all radial direction. This procedure enables us to measure the characteristic length scale of such irregular crater shape. 

\begin{figure}
\begin{center}
\includegraphics[width=7.5cm]{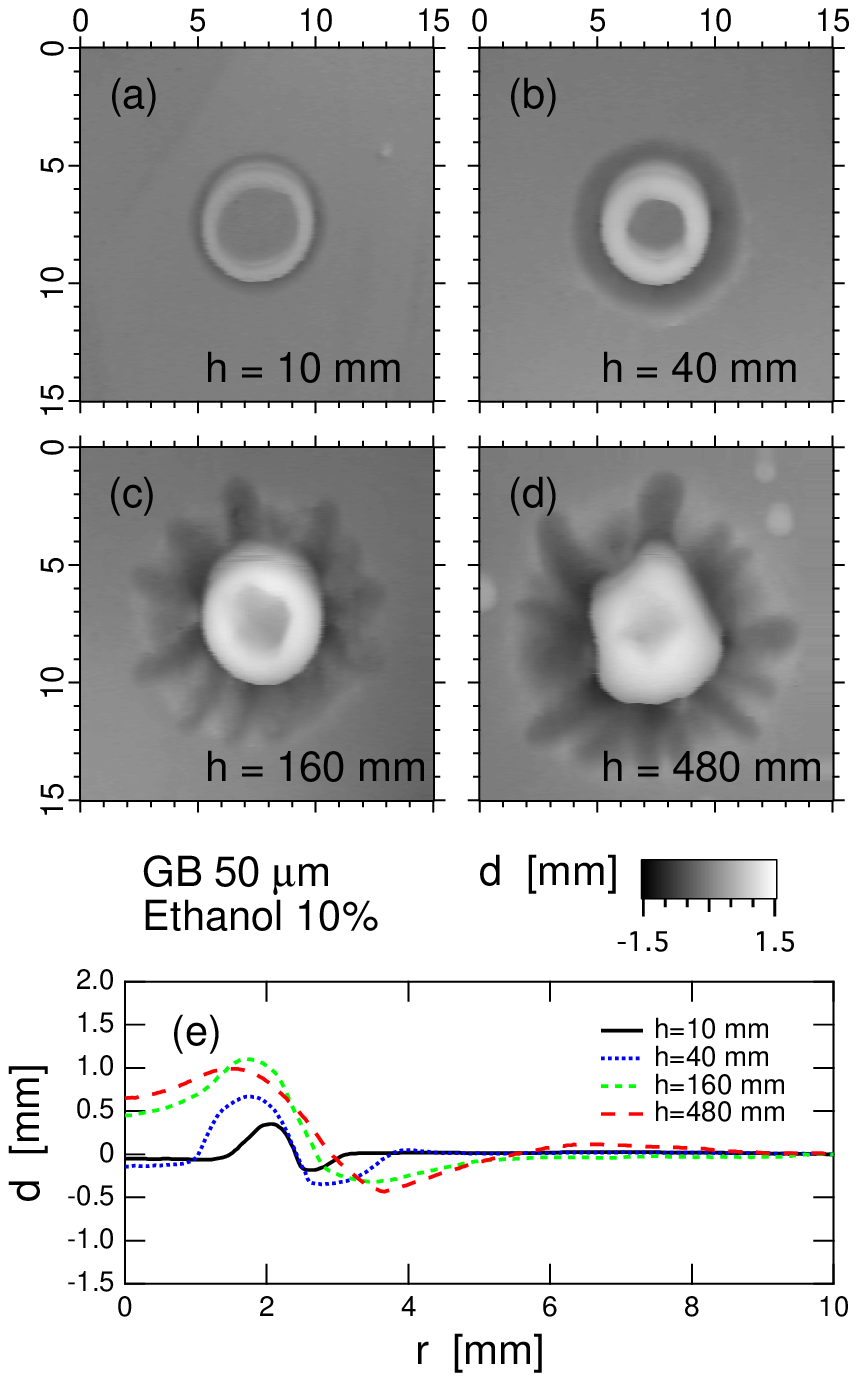}
\caption{Surface data of glass beads ($D_g=50$~$\mu$m) vs 10wt\% ethanol solution drop ($R_l=1.9$~mm). (a-d) Surface height maps are shown as gray scale images. (e) Corresponding radial depth functions $d(r)$ are computed from raw height maps (a-d). Crater shapes are similar to those in figure~\ref{fig:S3}. Drop fingering instability is enhanced due to the low surface tension. 
}
\label{fig:S4}
\end{center}
\end{figure}

In figure~\ref{fig:S4}, ethanol (10wt\% aqueous solution) drop impacts to the glass beads layer ($D_g=50$~$\mu$m) are shown. Roughly speaking, the crater structure in figure~\ref{fig:S4} is similar to that of figure~\ref{fig:S3}. Due to the low surface tension, fingering instability is enhanced and the remnant wavy rim structure is also emphasized~(figures~\ref{fig:S4}(c,d)). As mentioned before, pure ethanol ($100$wt\%) drop impact causes the drop fragmentation~(figures~\ref{fig:photos}(h,i)). Then the resultant crater size becomes small. This drop fragmentation effect is behind the scope of this paper. We focus only on the drop deformation and the drop-granular-interaction. Therefore, we quit measuring the characteristic length scale of the pure ethanol drop impact. Namely, we use only the $10$wt\% ethanol drop impacts data. The surface tension of 10wt\% ethanol solution is $47$~mN/m~(table~\ref{tab:fluids}). This value is about 65\% of the pure water's surface tension. This means that it is hard to vary the surface tension widely without drop fragmentation. If the surface tension is a little lower than water, drop fragmentation relatively easily occurs. 

\begin{figure}
\begin{center}
\includegraphics[width=7.5cm]{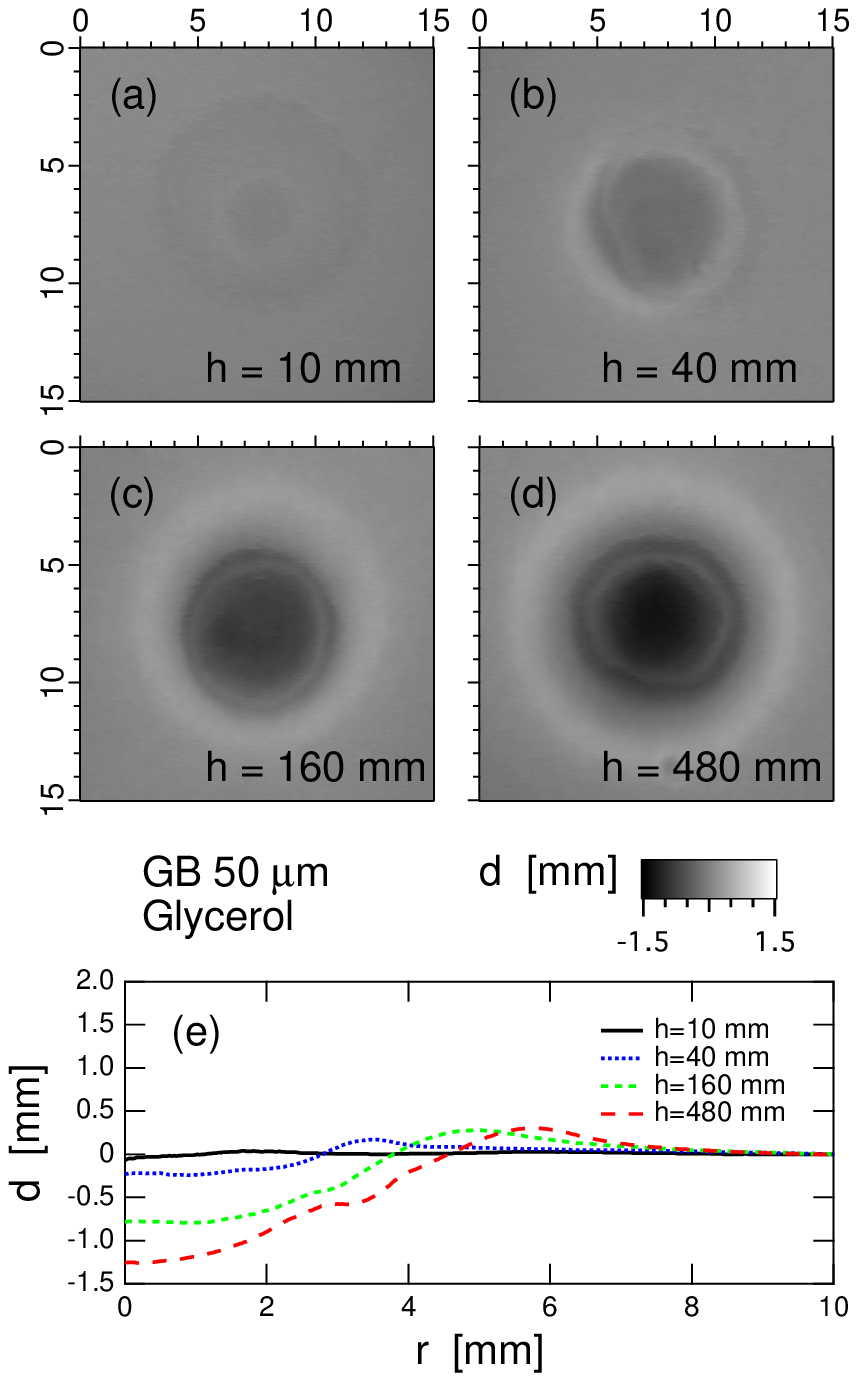}
\caption{Surface data of glass beads ($D_g=50$~$\mu$m) vs pure glycerol ($100$wt\%) drop ($R_l=1.9$~mm). (a-d) Surface height maps are shown as gray scale images. (e) Corresponding radial depth functions $d(r)$ are computed from raw height maps (a-d). All crater shapes look smoother than other cases.  
}
\label{fig:S5}
\end{center}
\end{figure}

Finally, we show the glycerol drop impacts to the glass beads ($D_g=50$~$\mu$m), in figure~\ref{fig:S5}. In this case, crater shapes are smooth. At low impact speed, almost the flat surface remains. At high impact speed, a simple concave crater appears with a thin inner ring. Consequently, corresponding radial depth functions $d(r)$ are calm. This calm shape is certainly caused by the viscosity effect. Viscous resistance prevents the drop from absorbing surrounding grains. And the drop penetrates into the granular layer very slowly. This process makes smooth concave crater shape. To characterize this viscous effect quantitatively, we have measured the characteristic time scales. The relation between time scales and viscosity is discussed in the next section. 

\section{Analyses}\label{sec:Analyses}
\subsection{Time scale analysis}\label{sec:time_scale_analysis}
We define two characteristic time scales $t_b$ and $t_p$. When the drop hits the granular layer, the drop often rebounds and exhibits the dumping oscillation of the drop deformation. $t_b$ is the settling time of this oscillation. After $t=t_b$, the drop penetrates into the granular layer. $t_p$ characterizes this slow penetration. It is the duration from the impact moment to the complete penetration of the drop. In almost all data, $t_b$ and $t_p$ were measured by eye using high speed video data. In some cases, however, $t_p$ is too long ($t_p >100$~s) to be measured by high speed video data. In such cases, the impacts were taken also with a 30 fps video camera, and $t_p$ was measured based on this lower speed video data. 

\begin{figure}
\begin{center}
\includegraphics[width=7.5cm]{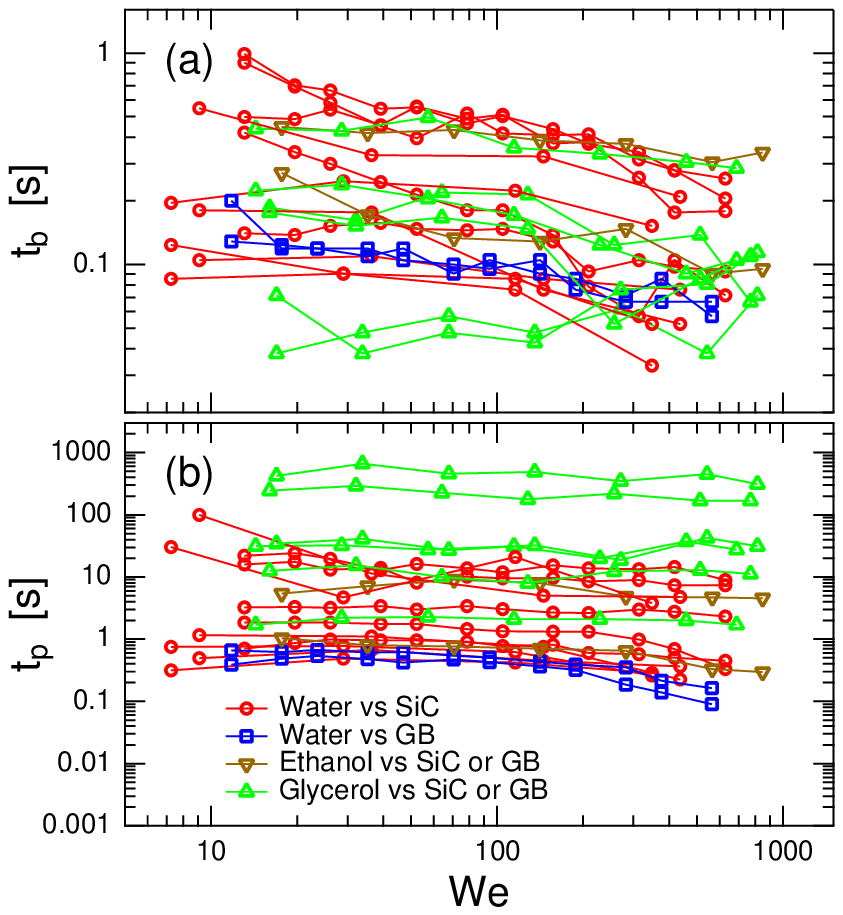}
\caption{Time scale data. (a) Drop oscillation settling time $t_b$ and (b) drop penetration time $t_p$ are shown as functions of $We$. The identical experimental conditions data (except $h$) are connected. In almost all situations, $t_p$ is roughly independent of $We$ over two orders of magnitude. While $t_b$ slightly depends on $We$, the dependence is very weak. Thus we regard it as roughly constant. Each point corresponds to a single impact experiment.
}
\label{fig:t_We}
\end{center}
\end{figure}

In figure~\ref{fig:t_We}, the measured time scales $t_b$ and $t_p$ are shown as functions of $We$. $We$ is the Weber number defined as follows, 
\begin{equation}
We = \frac{2 \rho_l R_l v^2}{\gamma}.
\label{eq:We}
\end{equation}
Where $\gamma$ is the surface tension of liquid drop. In figure~\ref{fig:t_We}, the same experimental conditions data (except the free fall height) are connected. Both of $t_b$ and $t_p$ vary in broad ranges; $0.01 \leq t_b \leq 1$ and $0.1 \leq t_p \leq 1000$ (s). Nevertheless, it seems that the time scales are almost independent of $We$, in which the principal variable is the impact speed. The time scales actually depend on viscosity. Moreover, the time scales depend on the granular media. This granular dependence seems to come from wettability difference and surface roughness.

\begin{figure}
\begin{center}
\includegraphics[width=7.5cm]{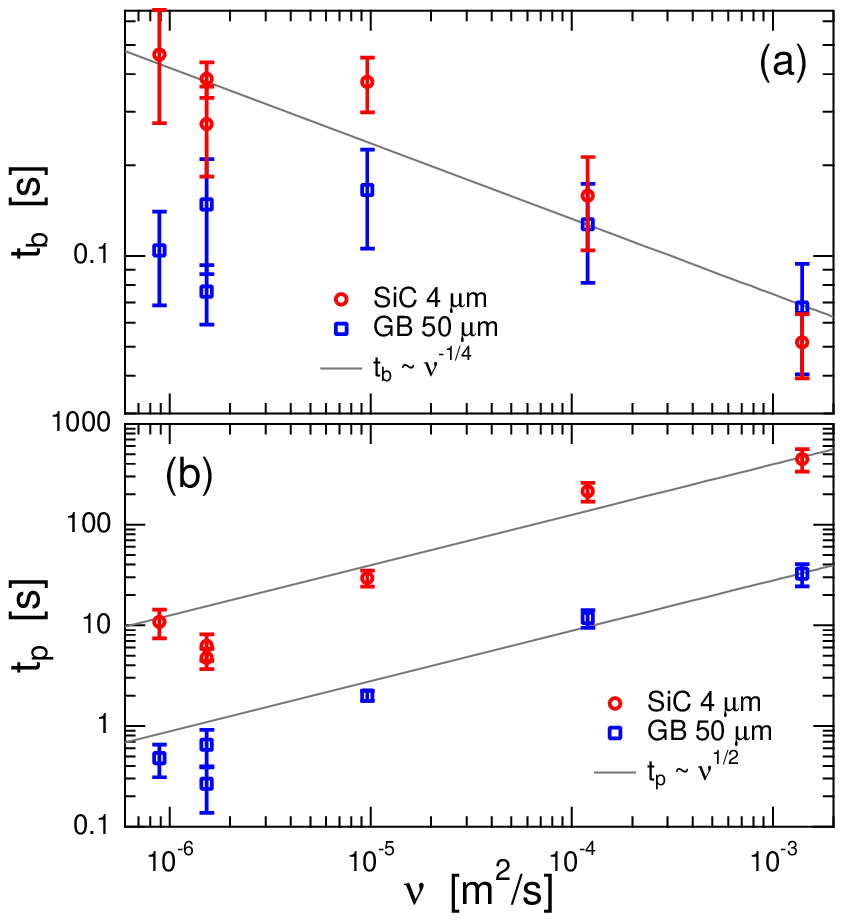}
\caption{Averaged time scales (a) $t_b$ and (b) $t_p$ vs kinematic viscosity $\nu$. $t_b$ for SiC seems to decrease as $t_b \sim \nu^{-1/4}$. Glass beads $t_b$ does not show obvious trend, particularly in low viscosity regime. $t_p$ for both cases show the scaling $t_p \sim \nu^{1/2}$. The $t_p$ of glass beads is much shorter than that of SiC due to its hydrophilicity.
 }
\label{fig:t_nu}
\end{center}
\end{figure}

In figure~\ref{fig:t_nu}, we plot average $t_b$ and $t_p$ vs kinematic viscosity $\nu$ for SiC ($D_g=4$~$\mu$m) and glass beads ($D_g=50$~$\mu$m). Each data point is calculated from the average of different free fall height results. As seen in figure~\ref{fig:t_nu}(a), $t_b$ of SiC is a decreasing function of $\nu$, while $t_b$ of glass beads may look almost constant. As shown in figures~\ref{fig:photos}(j,k), viscosity prevents the large deformation of liquid drop. Besides, the viscosity represents the dissipation rate for the dumping oscillation of drop deformation. As a result, $t_b$ shows decreasing trend by viscosity. The gray line in figure~\ref{fig:t_nu}(a) represents a tentative scaling $t_b \sim \nu^{-1/4}$.

In figure~\ref{fig:t_nu}(b), the penetration time $t_p$ is plotted as a function of $\nu$. Both of SiC and glass beads data show similar increasing trend. The scaling relation we obtain is approximately written as, 
\begin{equation}
t_p \sim \nu^{1/2}.
\label{eq:tp_nu}
\end{equation}
According to the Washburn and related recent result, the penetration time of liquid into porous media should be proportional to the viscosity, $t_p \sim \nu$~(\cite{Washburn1921,Hapgood2002}). However, scaling~(\ref{eq:tp_nu}) is quite different from their prediction. Although the granular layer is a certain kind of porous media, it is deformable in the current experiment. That might be a reason of different scaling exponent. The penetration time scale $t_p$ strongly depends on granular media as well. The $t_p$ of SiC is about one order of magnitude larger than that of glass beads. This implies that the SiC is much more hydrophobic than glass beads, as mentioned before.

\subsection{Length scale analysis}\label{sec:length_scale_analysis}
Some characteristic length scales are discussed in this section. Although the crater shapes are not very simple as shown in figures~\ref{fig:photos} and \ref{fig:S1}-\ref{fig:S5}, we can calculate some relevant length scales using $d(r)$. 

\begin{figure}
\begin{center}
\includegraphics[width=7.5cm]{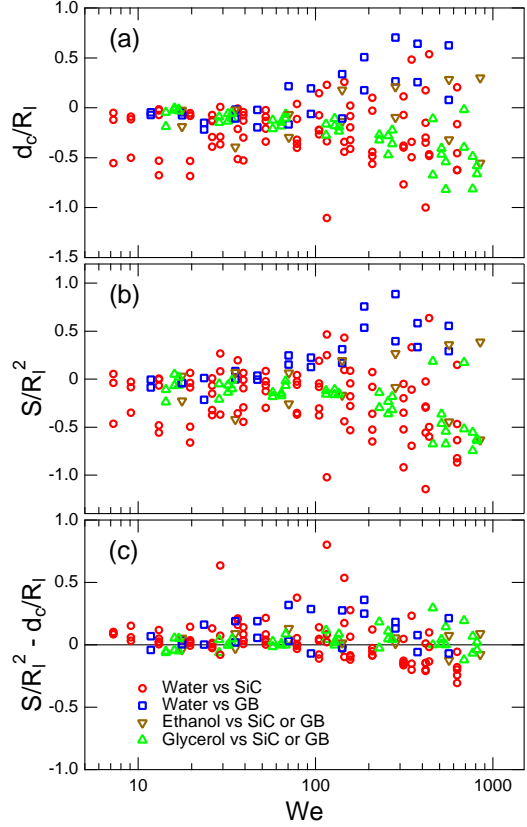}
\caption{(a) Normalized crater depth at the center $d_{c}/R_l$ and (b) normalized integrated area $S/{R_l}^2$ vs $We$. The global trends of $d_{c}/R_l$ and $S/{R_l}^2$ are very similar. (c) Subtraction of these two quantities is shown vs $We$. The relation $S/{R_l}^2 - d_{c}/R_l \simeq 0$ is roughly satisfied. This relation holds both for concave and convex crater shapes. 
}
\label{fig:d_We}
\end{center}
\end{figure}

First, the vertical length scale is discussed. We define the central depth at $r=0$ as $d_{c}=d(0)$. The normalized central depth $d_{c}/R_l$ is shown in figure~\ref{fig:d_We}(a). In order to characterize this data, we introduce a normalized integration of $d(r)$ as $S/{R_l}^2=(1/{R_l}^2)\int_0^{\infty} d(r) dr$. The $S/{R_l}^2$ indicates the compression or dilatation rate of the impacted granular layer. For usual solid projectile impact, $S/{R_l}^2$ should be negative. In other words, we expect that the impact always compresses the granular layer. In the drop-granular-impact, however, $S/{R_l}^2$ can be positive as shown in figure~\ref{fig:d_We}(b). When the drop hits the granular layer, it expands and absorbs surrounding grains. Then it penetrates into the granular layer with very gentle sedimentation of grains. This process creates a very sparse layer. That is a reason why the dilatation is possible by the drop-granular-impact. This dilatation can be observed especially for the heavier granular target cases. In such cases, the compression by the impact is not so remarkable.

We find that these two normalized values ($d_{c}/R_l$ and $S/{R_l}^2$) behave very similarly. To show this similarity, we plot $S/{R_l}^2 - d_{c}/R_l$ vs $We$ in figure~\ref{fig:d_We}(c). While the data scatter a little, most of all data are around $0$. This implies that the crater shape obeys a simple relation, 
\begin{equation}
S \simeq d_{c} R_l.
\label{eq:SdR}
\end{equation} 
Note that this relation holds both for concave (compressed) and convex (dilated) crater shapes. Equation~(\ref{eq:SdR}) means that the rate of compression or dilatation by the impact can be expressed only by $d_{c}$ and liquid drop radius $R_l$. Therefore, we regard $d_{c}$ as a representative vertical length scale for the drop-granular-impact. Honestly, the $d_{c}$ value itself cannot be predicted from the experimental conditions. It does not show simple $We$ dependence as shown in figure~\ref{fig:d_We}(a). The detail analysis to predict $d_{c}$ behavior is an open problem.

Next, the horizontal length scale is discussed. Once we get the $d(r)$ function, it is easy to calculate the crater radius $R$. $R$ is defined by the outer rim radius. In the previous report~(\cite{Katsuragi2010_1}), we found that the crater radius exhibits the scaling,
\begin{equation}
\frac{R}{R_l} \sim \left( \frac{\rho_g}{\rho_l} \right)We^{1/4}.
\label{eq:R_We}
\end{equation}
In solid projectile granular impacts, similar $1/4$ scaling has been reported~(\cite{Amato1998,Uehara2003,Walsh2003}). However, the density ratio dependence is opposite to the drop-granular-impacts result. On the other hand, $We^{1/4}$ scaling for the water drop deformation has been reported in the literatures~(\cite{Okumura2003,Clanet2004,Biance2006}). They have studied a water drop impact onto a superhydrophobic substrate~(\cite{Richard2000}). Since the crater radius seems to be determined by the drop deformation degree, the drop deformation scaling is more plausible than the analogy of solid projectile impacts, to explain the scaling exponent $1/4$~(\cite{Katsuragi2010_1}). 

\begin{figure}
\begin{center}
\includegraphics[width=7.5cm]{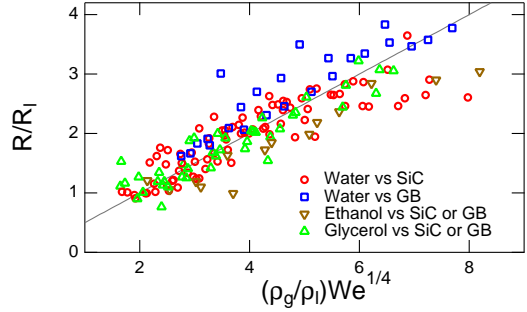}
\caption{Scaling plot of the crater radius, $R/R_l \sim (\rho_g/\rho_l){We}^{1/4}$. While this plot includes different viscosity results, obtained $R$ data seem to be independent of viscosity. The relevant effective parameter to control the crater radius is $R_l(\rho_g/\rho_l){We}^{1/4}$.
}
\label{fig:R_We14}
\end{center}
\end{figure}

In this study, we varied target grains (size and materials), drop surface tension, and drop viscosity. Using these data, we test the universality of the scaling~(\ref{eq:R_We}). In figure~\ref{fig:R_We14}, normalized crater radius $R/R_l$ is plotted as a function of $(\rho_g/\rho_l)We^{1/4}$. While the data are noisy, the global trend follows the scaling~(\ref{eq:R_We}). We can reduce the surface tension only to $65$\% of pure water, using $10$wt\% ethanol solution. The surface tension variation range is actually very limited. The scaling relation~(\ref{eq:R_We}) is robust at least in this surface tension variation range. Figure~\ref{fig:R_We14} includes all different viscosity data. Maybe the glycerol impact data are under the scaling line for high $We$ regime. the size dependence on viscosity might slightly exist. However, the $R/R_l$ basically looks independent of the viscosity. Although the viscosity was varied over three orders of magnitude, the crater radius is still independent of viscosity. This result is surprising since the drop deformation depends on viscosity (see e.g., figures~\ref{fig:photos}(d,k)). 

\section{Discussion}\label{sec:discussion}
To vary the crater size, density ratio control is much better than surface tension control. Low surface tension drops easily fragment to small drops by the impact. Moreover, the surface tension affects the crater radius as $\gamma^{-1/4}$ while the density ratio affects as $\rho_g/\rho_l^{3/4}$. The crater radius is independent of $\nu$ and is scaled to the impact speed as $v^{1/2}$. Thus $\nu$ or $v$ based control is not so efficient. The density ratio is a quite important parameter to control the crater radius. This might mean that the momentum transfer between granular layer and projectile is still the most important even for the deformable projectile impact. If we assume the momentum transfer as $\rho_l R_l^3 v^2 / d_1 \sim \rho_g R_h R_l$ with $d_1 \sim R_l^3/R_h^2$, we obtain the relation $R_h/R_l \sim \rho_g/\rho_l$. Where $R_h$ is the horizontal length scale of the deformed drop.

In low viscous drop impact, surface tension governs the drop deformation and naturally satisfies (\ref{eq:R_We}). In high viscous drop impact, we expect that viscosity effect must govern the drop deformation. In figures~\ref{fig:photos}(j,k), the drop deformation is suppressed by the viscosity. However, the resultant crater radius obeys the scaling~(\ref{eq:R_We}) which is independent of the viscosity. As long as we observe high speed video data, the deformation rate is quite different between water drop and glycerol drop impacts. Nevertheless, the normalized crater radius data cannot distinguish them. This means that the data collapse mainly comes from the density ratio effect. The density of viscous drop $\rho_l$ is relatively large (table~\ref{tab:fluids}). Thus the density ratio becomes small and hence drop deformation and $R$ also become small. As described above, the density ratio significantly affects the crater radius. Therefore, the scaling~(\ref{eq:R_We}) is recovered even for very viscous drop impacts.

Contrastively, the characteristic time scales depend on viscosity as $t_b \sim \nu^{-1/4}$ (only for SiC) and $t_p \sim \nu^{1/2}$ (both for SiC and glass beads). They are almost independent of the impact speed. The $t_p$ scaling is incompatible with traditional Washburn law. A new theoretical model might be necessary to solve this problem.

Since glass is very hydrophilic, the impactor drop can collect a lot of glass beads much easier than SiC grains. As a result, the collected many grains produce large inner ring or bump for the glass beads impacts. On the other hand, viscosity affects as a resistance against the grains absorption. That is why the viscous drop impact cannot make clear inner ring or bump. 

Grain size also affects the surface hydrophilicity. And the surface hydrophilicity influences the fingering and fragmentation instability of the impacted drop. We have not measured the detail of the wettability such as the contact angle. We have used only two kinds of grains, SiC or glass. Thus the discussion on hydrophilicity is still qualitative. More detail experiments and analyses are required to fully understand the hydrophilicity effect. It is the important problem open to future.

The origin of inner ring formation might be drop interior flow effect, such as the capillary flow~(\cite{Deegan1997,Deegan2000}). Maybe the fluid flow on the drop surface is measureable using absorbed grains as tracers. By \cite{Clanet2004}, flow measurement has been carried out and the vortical flow has been found for the impact between a liquid drop and a hard wall. Flow visualization in the drop is one of the future problems.

Characterization of drop fingering instability has been also studied~(\cite{Marmanis1996,Throddsen1998,Bhola1999}). They have tried to identify the boundary between fingering instability and smooth expansion. We observed fingering instability in the drop-granular-impact as well. Detail study of this instability is also an interesting future problem. 

Some dimensionless numbers made by combinations of Weber and Reynolds numbers have been proposed to take into account viscosity effect for the drop impact analysis. Marmanis and Throddsen have derived the impact Reynolds number as $I=We^{1/4}Re^{1/2}$~(\cite{Marmanis1996}). Clanet et al. have proposed $Re^{1/5}$ scaling for the impact drop deformation~(\cite{Clanet2004}). We tried applying these scaling relations to our data. Unfortunately, they do not collapse our data. The simplest one (\ref{eq:R_We}) brings the best result. More detailed investigation about the viscosity effect might be required to fully understand the drop-granular-impact dynamics.

\section{Conclusion}\label{sec:conclusion}
We performed the drop-granular-impact experiment using various liquid drops and granular layers. We found that the morphology of these impacts is basically similar to the previously reported result~(\cite{Katsuragi2010_1}). Grains hydrophilicity affects the impact dynamics and crater morphology. Glass beads layer cannot make clear sink-type crater. Small glass beads layer has large surface strength due to the capillary bridge based cohesive force. This strong granular layer results in the drop deposition rather than the cratering. Low surface tension drop impact induces fingering instability and fragmentation of the drop. In that case, underlying physical mechanisms are different from the single drop impact cratering. Thus we removed the data from quantitative analyses. Using the relevant data, we analyzed the characteristic time and length scales. We found that the characteristic time scales are almost independent of impact speed. An empirical scaling for the penetration time scale is obtained as, $t_p \sim \nu^{1/2}$. Next, we found the simple morphological relation between the impact compression or dilatation and the crater depth as $S \simeq d_{c}R_l$. Finally, the previously proposed crater radius scaling $R/R_l \sim (\rho_g/\rho_l)We^{1/4}$ was confirmed in a broad range of the parameter space. Particularly, it is surprising that the crater radius is independent of viscosity.

\begin{acknowledgments}
The author thanks T. Yamaguchi and K. Okumura for useful discussion. This research has been supported by the Japanese Ministry of Education, Culture, Sports, Science and Technology (MEXT), Grant-in-Aid for Young Scientists, No.~21684021.
\end{acknowledgments}

\bibliography{katsuragi_drop}

\end{document}